# Computing the Performance of A New Adaptive Sampling Algorithm Based on The Gittins Index in Experiments with Exponential Rewards

James K. He [1] [2], Sofía S. Villar [1], and Lida Mavrogonatou [1]

[1] University of Cambridge, Cambridge CB2 1TN, UK, kh672@cantab.ac.uk
[2] Yonder Technology Limited, London EC1V 9HX, UK

**Abstract.** Designing experiments often requires balancing between *learning* about the true treatment effects and *earning* from allocating more samples to the superior treatment. While optimal algorithms for the Multi-Armed Bandit Problem (MABP) provide allocation policies that optimally balance *learning* and *earning*, they tend to be computationally expensive. The Gittins Index (GI) is a solution to the MABP that can simultaneously attain optimality and computationally efficiency goals, and it has been recently used in experiments with Bernoulli and Gaussian rewards. For the first time, we present a modification of the GI rule that can be used in experiments with exponentially-distributed rewards. We report its performance in simulated 2-armed and 3-armed experiments. Compared to traditional non-adaptive designs, our novel GI modified design shows operating characteristics comparable in *learning* (e.g. statistical power) but substantially better in *earning* (e.g. direct benefits). This illustrates the potential that designs using a GI approach to allocate participants have to improve participant benefits, increase efficiencies, and reduce experimental costs in adaptive multi-armed experiments with exponential rewards.

## 1 Background

### 1.1 Response-Adaptive Experiments

The goal of conducting experiments is to find the true differences between treatment arms in order to provide real benefits to the wider population. However, experiments may also have other objectives, such as directly benefiting participants and improving experimental efficiencies. These objectives are important, but they can often be conflicting. For example, allocating more participants to the potentially superior arm may provide better benefits during the experiment, but it reduces the sample size available to the potentially inferior arm and lowers the overall statistical power. Balancing the *learning* objective of higher statistical power and lower bias with the *earning* objective of greater direct benefits to participants has thus been an intrinsic challenge in experiment design [1].

| Learning | Earning |
|---|---|
| Which advertisement is better | Maximise clicks during the trial |
| How to best promote referrals | Increase referrals during the trial |
| Which UI do users prefer | Minimise churns from bad experiences |
| What content interest the student | Keep the student engaged |

**Table 1.** Learning and Earning Objectives in Different Experiments

Traditional experiments are generally designed to optimise the *learning* component of the experimental objectives. To do this, most experiments randomly allocate participants to different treatment arms with equal probability fixed throughout the experiment, a design known as Randomised Control Trial (RCT) in academia and A/B Testing in industry. As such, every treatment arm has roughly the same number of participants, optimising statistical power when making comparisons [2].

However, this Equal Randomisation (ER) design largely ignores the *earning* objective, since it allocates the same number of participants to potentially inferior arms as well as superior ones. Thus, ER design decreases the direct outcomes or participant benefits during the experiment. This is a widespread challenge for experiments in general (see **Table 1.** for examples), and can especially become an ethical dilemma in clinical trials. For example, in trials for fatal or rare disease treatments, participants can die if allocated to the inferior arm, or they may be significant proportions of all patients for a rare disease. In these cases, the need for directly *earning* participant outcomes during the trials arguably outweighs the future outcomes granted by the *learning* achieved during the trials.

Response-adaptive experiments [3] [4] [5] are proposed to solve this limitation of traditional ER experiment designs. In general, adaptive designs dynamically adjust the probability for each new participant to be allocated into each arm based on existing observations. By allocating more participants to the superior arm, adaptive experiments achieve better direct participant benefits compared to ER designs. Having less participants in the inferior arms, however, reduces the statistical power [6] [7] [9] when making comparisons between different arms. It is consequently an important theme in response-adaptive designs to balance the *learning-earning* trade-off, where it would be favourable to have a slight reduction in statistical power if it allows substantial improvement in direct participant benefits during the experiments.

### 1.2    Multi-Armed Bandit Models

Many sample allocation strategies have been proposed to solve the optimal *learning*-v.-*earning* dilemma described above (e.g., [11] [12]), one such formal solution is to model adaptive experiments as Multi-Armed Bandit Problems (MABP) [3] [13] [15]: A 1-armed bandit is a process that produces a random payoff of unknown distribution when activated (e.g., a slot machine, i.e., a bandit); consider an *M*-armed bandit where some arms may have more favourable payoff distribution than others, and that only one of the *M* arms can be observed at each time, how to sequentially observe each arm to yield the maximum total payoff? This requires choosing the best-performing arm as much as possible, yet still observing other arms enough to discover which arm is truly best-performing.

More formally, consider successive plays *i*(*t*) of arm *m*, with (*m* = 1,2,...,*M*), yield i.i.d. reward sequences $\{Y_i^m\}^\infty_i$ which we assume to be exponentially distributed with an unknown parameter $\lambda_m$. The MABP in this case is thus a problem of finding the optimal sampling policy among the arms over time, *i*<sup>∗</sup>(*t*), defined as:

$$i^*(t) = \arg\max_{i(t) \in I} E_0^{i(t)} \sum_{t=1}^{\infty} Y_{i(t)} d^t \tag{1}$$

where *I* includes all sampling policies *i*(*t*) that sample only 1 arm at a time *t*, 0 represents the initial information on each arm before sampling any observation and $d \in [0,1)$ is a discount factor.

In the context of adaptive experiment designs, the MABP may be reformulated as: with *M* number of treatment arms that produce participant outcomes of unknown distributions, where some arms may produce more favourable outcomes than others, and that each participant can only be allocated to one of the *M* arms, how to sequentially allocate participants to arms based on the initial information and the observed outcomes so far to yield the maximum participant outcomes? In other words, how to allocate as many participants as possible to the currently best-performing arm, yet exploring other arms enough to determine the true best-performing arm?

Therefore, it is clear that an optimal solution to the MABP can provide an optimal adaptive experiment design that maximises participant outcomes, upon which constraints can be added to balance statistical power for the *learning* objective. However, many solutions to the MABP, such as the dynamic programming solution [14], can become computationally intractable as the number of arms *M* increases, especially in continuous state space such as with an exponential reward process [7] [8]. In the MABP literature, index-based solutions (e.g., [17] [18]) have thus been proposed as computationally efficient and near-optimal solutions to the MABP. The present work focuses on one of the most prominent amongst such index-based solutions: the Gittins Index [16] [17].

### 1.3    Previous Work

Formal definitions of the Gittins Index (GI) were developed in the 70's [16] [17] and they have been more recently described by [6]. We refer the reader to that survey paper for a formal framework of the index approach. The majority of the work assessing the GI policies is in the context of a binary reward (as is the case for the work reviewed in [6]). Previously, GI-based designs have also been shown to perform well in simulated experiments with continuous normally-distributed rewards [7]. Compared to traditional ER designs, these simulation studies found that GI-based designs generally perform slightly worse in statistical power and accuracy of the estimator, but perform substantially better in terms of direct participant benefits during the experiments.

However, to the best of our knowledge, with the exception of one recent work [7], there is no literature reporting the performance of GI-based policies in experiments with exponentially-distributed outcomes. This presents a significant limitation in the current response-adaptive experiment design literature. In clinical trials, for example, exponential models have been shown to potentially improve trial efficiencies by 35% and reduce required sample sizes by 28% [19]. Many nonclinical experiments may also have exponentially-distributed outcomes and are therefore subject to the same limitation, such as those where a small number of observations have very large values (e.g., income, engagement time, referral frequencies). Designing response-adaptive experiments with exponential rewards has the potential to significantly improve direct participant benefits, increase experimental efficiencies, and reduce costs for experiments in a wide range of contexts, from clinical trials and A/B testing for web design, to personalised e-learning and media content.

Here, we present a new variant of a GI-based adaptive design to experiments with exponential rewards. We report the operating characteristics of our modified GI designs in multiple simulations, comparing them with a balanced sampling approach (referred to as ER, also known as RCT or A/B testing). Since GI is still computationally feasible when there are a larger number of treatment arms ($M$), our modified GI designs have the potential to improve direct participant benefits by facilitating low-cost designs of experiments with multiple arms.

## 2    Method

### 2.1    GI in Exponential Reward Process

The Gittins Index rule is a policy for making optimal sequential selection of arms in an infinite horizon discounted MABP: each arm can be assigned a GI, and in order to maximise the expected total outcome one needs to always select the arm with the highest GI. GI's are dependent upon different reward distributions, different scale parameters, a discount factor value and different numbers of observations. Derivations and calculations of GI's have been explained in [17].

Following the instructions laid out in [7], this work calculates the GI for each arm at each new observation from the values tabulated in [17]. Specifically, the tables record index values $v(n,d,1)$ for each number of observations $n$ at a reward discount rate of $d$ under an exponential reward process with mean $\mu = \lambda = 1$. A version of this table is attached in the Supplementary Materials, in which values originally un-tabulated in [17] are interpolated as per instructed by the original work. R scripts for the interpolation and the allocation algorithm are also available in the Supplementary Materials.

Following [7], we adopt a Bayesian framework in which a prior distribution represents our initial beliefs on the unknown model parameters; upon observation of new data, the prior distribution is updated into the posterior distribution which incorporates this new knowledge. Specifically, for every treatment arm, a prior GI is assigned with an implicit prior where all arms have equal mean at $n = 2$ (analogous to having 2 participants' worth of information as prior beliefs, see supplementary methods in [7] for details). Since theorem 7.11 in [17] states that new indices under different means can be calculated by $v(n,d,\mu) = \mu v(n,d,1)$, we calculate a new GI after each new observation from multiplying the posterior mean to the tabulated index value corresponding to the number of observations and the desired discount rate. Since only one observation can be made on one arm at a time, when an arm is not selected, its GI remains the same until it is next selected by the algorithm to receive a posterior update.

Due to the nature of the GI policy to prefer allocating as many participants as possible to the superior treatment arm (if there is one), final allocations with the original GI rule will be highly imbalanced (with cases of arms having no samples being possible). To avoid this, we introduce a constraint factor $k$ such that at least $\frac{1}{k}$ of the $t$ allocated participants are in each of the $M$ treatment arms. For example, when $k = 5$, the algorithm first checks if each treatment arm has received at least $\frac{t}{5}$ of the $t$ allocated participants (i.e., if $t = 10$ participants have been allocated, does each arm have at least 2 observations). If any arm have less than $\frac{1}{5}$ of the allocated participants, the algorithm will allocate the next participant to the arm with the least observation until the minimum observations per arm is met, in which case the algorithm follows the GI policy by allocating the next participant to the arm with the highest GI. In other words, this modification makes sure that a minimal proportion of the sample is allocated to each arm, but only interferes with the GI policy when any arm is severely neglected. We shall refer to this proposed GI-based modified implementation with a constraint factor as "the modified GI".

Theoretically, $k$ can take values between $M$ and $N$ where $M$ is the number of arms in the experiment (including control) and $N$ is the total number of participants expected in enroll in the experiments. When *k* = *M*, the allocation algorithm is constraint to the equivalence of ER designs; when *k* = *N*, only 1 participant is required to be allocated to each arm, allowing for the algorithm to potentially allocate all the remaining *N* – *M* participants to the superior arm. Since the latter optimal case is detrimental to statistical power (having only one observation is not a good way to learn), we conservatively set the upper bound of $k = \frac{N}{M}$ in this work, thereby ensuring that at least *M* participants are allocated to each arm. In order to have meaningful comparisons of operating characteristics such as statistical power, we therefore do not include an unrestrained GI design and instead consider the modified GI with $k = \frac{N}{M}$ as the "near-optimal" GI algorithm. Although not explored in this work, factor *k* ∈ [*M*,*N*] can also be treated as an unknown parameter that is sought to be optimised.

## 2.2    Simulation for 2-Armed Experiments

For simulating 2-armed experiments with exponential rewards, this work followed [9] and simulated the experiments under different potential scenarios with an arbitrary range of parameters. In this hypothetical experiment, there are in total *N* = 100 participants, with the control arm yielding outcomes that follow an exponential distribution with mean $\mu_0$ = 0.5, and the experimental arm yielding outcomes that follow an exponential distribution with mean $\mu_1$ ∈ {0.1 : 0.9}. The arbitrary sample size of 100 is chosen so that the simulated experiments have sensible ranges of statistical powers. The experiment has null hypothesis $H_0$ : $\mu_0$ = $\mu_1$ and alternative hypothesis $H_1$ : $\mu_0$ ≠ $\mu_1$, and uses the statistic $-F_{(N_1, N_0)}(\frac{\mu_0}{\mu_1})$ [10] to test the hypothesis at a cutoff of $\alpha$ = 0.05.

First, 10000 experiments are simulated under $H_0$ to observe the Type I error (false-positive) rate by counting the number of these null experiments that returned significant results. Then, for each of the potential $\mu_1$ values, 10000 experiments are simulated to observe the statistical power by counting the number of true-positive results. Other operating characteristics, namely the proportion of participants allocated to the superior arm (if there is one), bias in the estimated $\mu$ from the experiment, and the total participant outcomes calculated as the sum of all observed outcomes in each arm, are also recorded. This entire process is then repeated for ER design and our modified GI designs with constraint factor *k* ∈{5,9,50}, where $GI_{k=50}$ is when *k* takes our upper bound of $\frac{N}{M}$ and referred to as "near-optimal".

## 2.3    Simulation for 3-Armed Experiments

To demonstrate the performance of the modified GI designs in experiments with multiple arms, we only simulate 3-armed experiments in addition, since operating characteristics for experiments with more than 3 arms become difficult to visualise. Set-ups for the 3-armed experiments are similar to the 2-armed case, with the differences that the allocation algorithm now records 3 Gittins Indices for each of the three arms, and that apart from varying $\mu_1 \in \{0.1 : 0.9\}$, we also vary $\mu_2 \in \{0.2 : 1.0\}$ to observe a wider ranges of parameter differences. For the same reason, we also set $\mu_0 = 0.4$ instead of 0.5. These 3-armed experiments have null hypothesis $H_0 : \mu_0 = \mu_1 = \mu_2$ and alternative hypotheses $H_{1A} : \mu_0 \neq \mu_1$, $H_{1B} : \mu_0 \neq \mu_2$, using the same test statistic as in Section 2.2 but with a different $\alpha$ cutoff after a Bonferroni correction for multiplicity to maintain a Family-Wise Type-I Error Rate of 0.05. Simulations for the 3-armed experiments are also the same as the 2-armed case, but only 5000 experiments are simulated for each potential $\mu_1$ and $\mu_2$ combinations to limit simulation runtime, and that the modified GI designs now have constraint factor $k \in \{5, 9, 33\}$ to reflect the change in the number of arms $M$.

Since two tests are performed in these 3-armed experiments, statistical power is calculated based on Family-Wise Error Rates, where an overall Type-I error is the percentage of experiments simulated under the null to falsely reject $H_0$ in favour of either alternatives, and Power is 1 minus the percentage of experiments simulated under the alternatives to falsely adopts $H_0$ in rejection of both alternatives. All other operating characteristics recorded for the 2-armed experiments are also included.

### 2.4    Operating Characteristics

To evaluate the performance of different experiment designs, four operating characteristics are computed. Since the present work aims to compare the *learning* and *earning* performance of different experiment designs, we pay specific attention on two operating characteristics linked to *learning*: 1) the statistical power (and hence the Type I error rate when under the null), and 2) the standard deviation of the estimate ($\sigma_{\text{Estimate}}$), as a measure of estimation accuracy; as well as two operating characteristics linked to *earning*: 3) the proportion of participants allocated to the superior arm ($\rho_{\text{superior}}$), and 4) the percentage increase from Expected Total Outcomes (% Increase in ETO).

The measure "% Increase in ETO" is calculated by taking a sum of all the participant outcomes in an experiment and comparing this observed total outcome to the expected total outcome when allocating participants under the ER design. For example, in a 2-armed case, the measure is calculated using the formula $(\sum x / (\frac{1}{2} n \mu_0 + \frac{1}{2} n \mu_1) - 1) \times 100$). The same four operating characteristics are also recorded for 3-armed experiments with the statistical power calculated from Family-Wise Error Rates as explained in the previous section.

## 3   Results

### 3.1   2-Armed Experiments

**Figure 1** summarises the simulated *learning* performances of the different experiment designs. On both panels, the horizontal axes represent the different values $\mu_1$ can take while $\mu_0$ remains fixed. The vertical axis on the left panel represents the statistical power, and that on the right panel represents the accuracy of the estimate measured by the estimate's standard deviation. Black dots represent operating characteristics from ER design; red, blue, and green dots represent operating characteristics from the modified GI designs with constraint factor *k* = 5,9,50, respectively.

Results for *learning* performance shows that, as expected, the ER design performs the best under all potential $\mu_1$ values in terms of power, and the GI(near-optimal) design performs the worst in terms of accuracy measured by standard deviation of the estimate. This is as expected since the modified GI designs attempts to balance *learning* and *earning* by attempting to allocate more participants to the superior arm, which inevitably decreases statistical power. When $\mu_1 = \mu_0 = 0.5$, the statistical power (in this case where the $H_0$ is true, it is the same as the Type I error rate) of all designs reduces to the preset $\alpha = 0.05$, as expected.

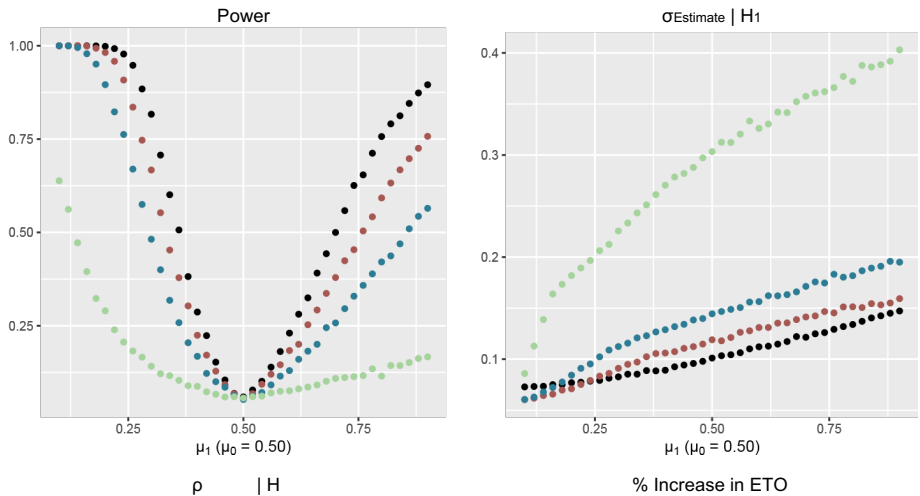

**Fig.1.** Operating Characteristics for Learning from ER and modified GI Designs in 2-Armed Simulated Experiments

Next, **Figure 2** summarises the simulated *earning* performances of different experiment designs. With the horizontal axes and different colours of dots representing the same as above in **Figure 1**, the vertical axis of the left panel represents the proportion of participants allocated to the superior treatment arm, and that of the right panel represents the percentage increase in Expected Total Outcome compared to ER expected total outcomes.

Both the *earning* operating characteristics show that the three modified GI designs substantially outperforms the ER design in terms of participant benefit improvement. When $\mu_1 = \mu_0 = 0.5$, all experiment designs allocate 50% of participants to each arm, as expected. However, while the ER design consistently allocates 50% of participants to each arm, the modified GI designs start allocating more participants to the superior arm as soon as a difference between $\mu_1$ and $\mu_0$ can be detected. The near-optimal GI($k$ = 50) design under-performs against other modified GI designs across most cases in terms of participant benefit. It also appears that the most constrained design GI($k$ = 5) performs the best and is only overtaken by the less-constrained GI($k$ = 9) after reaching its theoretical maxima (allocating 80% of all participants to the superior arm).

Notably, all four operating characteristics exhibit asymmetric behaviour, where the lower $\mu_1$ values receives higher statistical power, higher accuracy (lower $\sigma_{Estimate}$), and higher participant outcome measures, compared to the higher $\mu_1$ values that have the same distances from $\mu_0$. This is due to the nature of the exponential reward distribution where variance changes according to the mean; i.e., at smaller $\mu_1$ values, the rewards also have smaller variances, thus allowing better estimation (thereby improved power and accuracy) as well as better adaptive allocation by the modified GI algorithms.

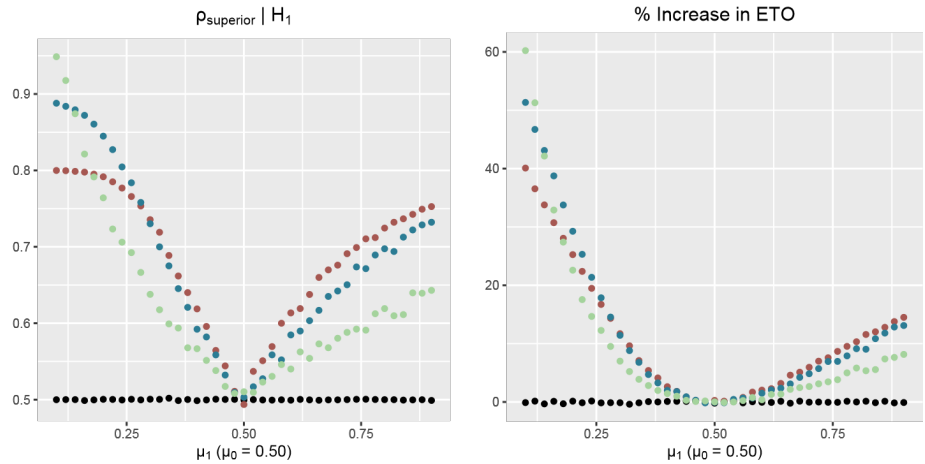

**Fig.2.** Operating Characteristics for Earning from ER and modified GI Designs in 2-Armed Simulated Experiments

## 3.2    3-Armed Experiments

Similar to results for 2-armed simulations, operating characteristics results for 3-armed simulations are also presented in figures that visualise the relationship between different $\mu$ values and the 4 operating characteristics. However, since we are varying both $\mu_1$ and $\mu_2$ in 3-armed experiment simulations, results are visualised in 3-dimensional figures with the *x* and *y* axes taking values of $\mu_1$ and $\mu_2$, and the *z* axis representing each of the 4 operating characteristics.

In **Figure 3**, results for statistical power (left panel) and accuracy (right panels) are presented. Green dots represent experiments with ER design, orange represents GI($k$ = 5), dark-blue represents GI($k$ = 9), and pink represents GI(near-optimal) (i.e., modified GI with $k$ = 33). Accuracy measured as standard deviation of the estimates are separately visualised for $\mu_1$ estimates (bottom-right) and $\mu_2$ estimates (top-right).

These results on *learning* performance shows that, as expected, the ER design outperforms all modified GI designs in terms of statistical power, and the GI(near-optimal) design under-performs all other designs in terms of accuracy by having substantially higher standard deviations of both estimates for $\mu_1$ and $\mu_2$. When $\mu_1$ = $\mu_0$ = 0.4 or $\mu_2$ = $\mu_0$ = 0.4, statistical powers of all 4 designs reduce to the Family-Wise $\alpha$ = 0.05, as expected. Notably, amongst the modified GI designs, GI($k$ = 5) and GI($k$ = 9) appear to perform nearly as well as the ER design in terms of both statistical power and accuracy in the two estimates.

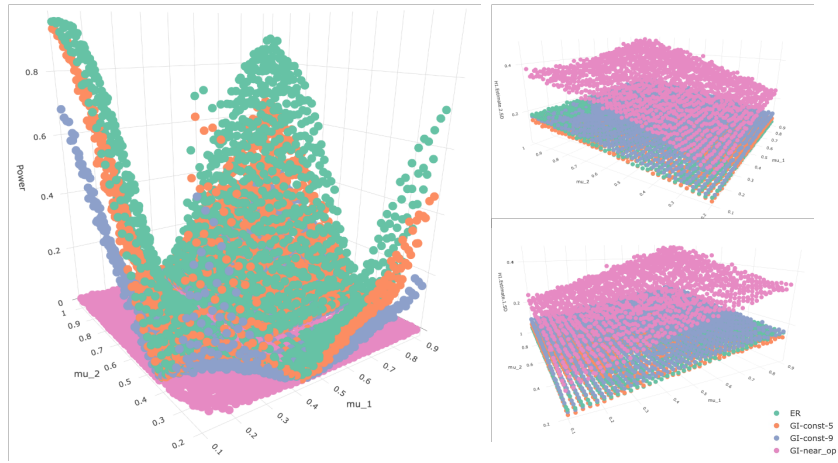

**Fig.3.** Operating Characteristics for Learning from ER and modified GI Designs in 3-Armed Simulated Experiments

Next, results for the *earning* performances of the different designs are summarised in **Figure 4**. Each colour has the same representations as in **Figure 3**. The vertical axis of the left panel represents the proportion of participants allocated to the superior arm, and that of the right panel represents the percentage increase in Expected Total Outcomes compared to the theoretical outcomes of the ER design. To aid the visual clarity, **Figure 4** has the $\mu_1$ and $\mu_2$ axes in descending scale instead of the ascending scale as appeared in **Figure 3**. This is simply a different rotational aspect of the same 3-dimensional visualisation, approaching from the higher ends of the two horizontal axes instead of from the lower ends.

Results on the *earning* performances of different designs show that the modified GI designs substantially outperforms ER in terms of both proportion allocated to the superior arm and percentage increase from Expected Total Outcome. As expected, when $\mu_1 = \mu_2 = \mu_0 = 0.4$, all designs allocate around $\frac{1}{3}$ of participants to each arm, and once there is a difference between the $\mu$ values, the modified GI designs allocate more participants to the superior arms while ER continues to maintain $\frac{1}{3}$ of participants in each arm. As a result, when $\mu_0$, $\mu_1$, and $\mu_2$ take extremely different values (e.g., $\mu_1 = 0.1$, $\mu_2 = 1.0$), the modified GI designs may yield a total participant benefit improvement of as much as 60%. Notably, GI($k = 9$) performs just as well as, if not better than, GI(near-optimal) design under most cases.

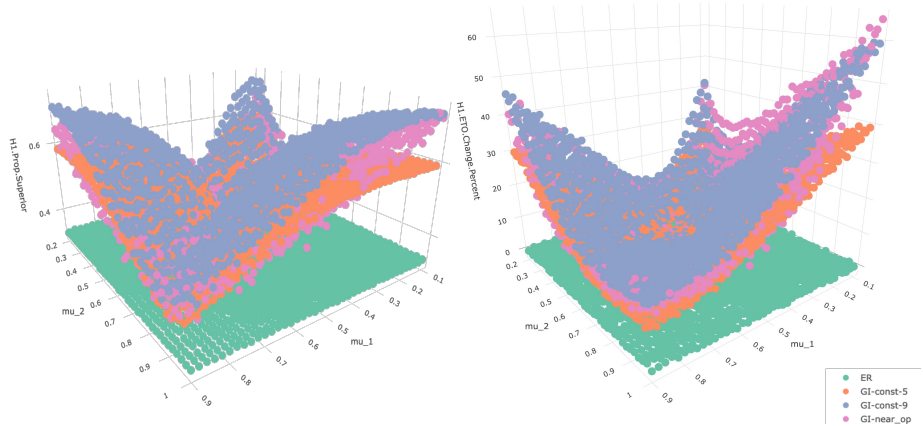

**Fig.4.** Operating Characteristics for Earning from ER and modified GI Designs in 2-Armed Simulated Experiments

The asymmetric behaviour exhibited in **Figure 3 & 4** can be attributed to the variance-dependence nature of the exponential reward process previously explained in 2-arm simulation results. In addition, our simulation setting that aims to observe a wider range of parameter differences, namely having $\mu_1$ and $\mu_2$ taking slightly difference ranges and having $\mu_0$ off-centred at $0.4$, also contributes to the apparent asymmetry in the results.

## 4     Discussion

The aim of experiments is often to balance various objectives, such as maximising statistical power and accuracy (the *learning* objective), and maximising direct participant benefit by allocating more participants to the superior treatment arm during the experiment (the *earning* objective). Gittins Index (GI), a computationally efficient and near-optimal solution to the Multi-Armed Bandit Problem, has the potential to guide adaptive experiment designs that balance these two objectives. Previous research has demonstrated that GI can be applied to experiments with Bernoulli and Gaussian rewards as a viable strategy for adaptive experiment designs. In this work, we extend GI to experiments with exponential rewards and present its performance, which has not been reported on before, to the best of our knowledge.

Based on simulations of 2-armed and 3-armed experiments with exponential rewards, we found that our modified GI designs perform stably in experiments with such reward distributions, extending the potential applications of the Gittins Index. In terms of operating characteristics, our modified GI designs perform slightly worse than the traditional ER design in terms of statistical power and accuracy, but substantially better in terms of participant benefits. The use of a constraint factor $k$ that regulates the minimum proportion of participants allocated to each arm appears to improve the learning performance of the modified GI designs without significantly reducing the participant benefit improvement.

Therefore, this work shows that modified GI adaptive experiment designs can be extended to experiments with exponentially-distributed outcomes, improving participant benefit without significantly reducing statistical power. This work also demonstrates that modified GI designs can be easily extended to 3-armed cases, which has the potential to improve efficiency and reduce costs by only requiring one control arm for testing the effects of multiple alternatives. The fact that the 3-armed extension was straightforward to implement and computationally feasible highlights a unique advantage of using index-based allocation strategies in multi-armed adaptive experiment designs.

Finally, the addition of a constraint factor $k$ in the modified GI allocation algorithm provides flexibility that allows experimenters to optimise the balance between the *learning* and *earning* objectives for their specific scenarios. Future research can build upon these results by extending GI-based designs to other outcome distributions. For the modified GI adaptive experiments with exponential end-points, future work can further investigate the behaviour of the GI designs under a wider range of constraint values $k$, numbers of arms $M$, and other relevant simulation parameters. Additionally, developing a general function for any $M$-armed adaptive experiment design would be a valuable contribution.

The potential applications of this work are numerous. For example, when the time spent viewing an advertisement is exponentially distributed, online advertisers can use the modified GI algorithms to experiment with different marketing messages, while minimising the exposure of inferior messages to a large proportion of the audience. Similarly, when a website-interaction metric is exponentially distributed, web designers can use the modified GI algorithms to experiment with different user interface designs, while minimising user churn due to bad experiences. Additionally, when content engagement outcomes are exponentially distributed, the modified GI algorithm can be used to personalise e-learning or media content by testing different content categories, while avoiding reductions in engagement from over-allocating inferior categories. Importantly, experimenters in all of these scenarios can adaptively test multiple options at once using an algorithm that is easy to implement and computationally efficient. As such, the modified GI algorithms may provide efficient solutions for adaptive A/B testing, content personalisation, and related reinforcement learning problems even in cases where the state space is continuous, as in our work.

## 5    Conclusion

Response-adaptive experiments aim to learn about the true effects of treatments and, at the same time, directly benefit participants by dynamically changing the allocation of participants to superior treatment arms during the experiment. One method for achieving this is to use the Gittins Index (GI), which is a computationally feasible solution to the Multi-Armed Bandit Problem, to guide the dynamic allocation of participants. In this work, we extend GI-based allocation algorithms to experiments with exponential rewards and report the performance of our modified GI algorithms compared to traditional equal-randomisation (ER, or A/B testing) designs. Our simulations of 2-armed and 3-armed experiments show that our modified GI designs can perform comparably to traditional designs in terms of statistical power, while also substantially improving participant benefit. By introducing an allocation constraint factor, we present an allocation algorithm that can be customised to meet specific needs. Overall, our results suggest that GI is a promising strategy for multi-armed adaptive experiment designs and has potential to substantially improve direct outcomes and reduce costs in clinical trials, UI testing, and content personalisation.

## A     Author Contributions

LM and SV defined the internship project proposal that led to this work; LM supervised the internship work. JKH and LM designed the simulation studies; JKH performed the simulations and analysed results; JKH drafted the manuscript; JKH, LM and SV contributed to the writing and editing of the manuscript.

## B     Acknowledgement

JKH thanks the University of Cambridge MRC Biostatistics Unit Summer Internship Program for the support and training that made this research possible. LM and SSV acknowledge funding and support from the UK Medical Research Council (MC UU 00002/15).

## C     Supplementary Materials

Table containing the interpolated Gittins Index values for the exponential reward process, as well as R-scripts of the interpolation and subsequent algorithm implementations, simulations, and visualisations, can be accessed through the GitHub repository: https://github.com/james-helium/gittins_adaptive_trials

## D     Data Access/Availability Statement

The results presented in this work were based solely on simulated data. The code used for generating the data and the subsequent analysis is available through the GitHub repository: https://github.com/james-helium/gittins_adaptive_trials

## E     Rights Retention Statement

For the purpose of open access, the author has applied a Creative Commons Attribution (CC BY) licence to any Author Accepted Manuscript version arising.